\documentclass[preprint,12pt]{elsarticle}

\usepackage{graphicx}
\usepackage{amssymb}

\usepackage{lineno}
\usepackage{listings}

\usepackage{mdwmath}
\usepackage{mdwtab}
\usepackage{multirow}
\usepackage{multicol}
\usepackage{array}
\usepackage{url}
\usepackage{booktabs}

\usepackage{enumitem}
\usepackage{xspace}
\usepackage[export]{adjustbox}

\newcommand{\ifReview}[1]{}

\usepackage{color,soul}
\soulregister\cite7 
\soulregister\citep7 
\soulregister\citet7 
\soulregister\ref7 
\soulregister\pageref7 

\newcolumntype{C}[1]{>{\centering\arraybackslash}p{#1}}
\newcolumntype{R}[1]{>{\raggedleft\arraybackslash}p{#1}}
\newcolumntype{L}[1]{>{\raggedright\arraybackslash}p{#1}}

\newcommand{\threePlusses}{\oplus\hspace{-0.5ex}\oplus\hspace{-0.5ex}\oplus}
\newcommand{\twoPlusses}{\oplus\oplus}
\newcommand{\onePlusses}{\oplus}

\usepackage{graphicx}
\usepackage{subcaption}  

\journal{Future Generation Computer Systems}

\begin{document}

\begin{frontmatter}


\title{uBaaS: A Unified Blockchain as a Service Platform}





\author[Data61]{Qinghua Lu}
\author[Data61]{Xiwei Xu\corref{cor}}
\author[upc]{Yue Liu}
\author[Data61]{Ingo Weber}
\author[Data61]{Liming Zhu}
\author[upc]{Weishan Zhang}

\address[Data61]{Data61, CSIRO, Sydney, Australia}
\address[upc]{College of Computer and Communication Engineering, China University of Petroleum (East China), Qingdao, China\\
qinghua.lu@data61.csiro.au, xiwei.xu@data61.csiro.au}
\cortext[cor]{Xiwei Xu is the corresponding author.}

\begin{abstract}
Blockchain is an innovative distributed ledger technology which has attracted a wide range of interests for building the next generation of applications to address lack-of-trust issues in business. Blockchain as a service (BaaS) is a promising solution to improve the productivity of blockchain application development. However, existing BaaS deployment solutions are mostly vendor-locked: they are either bound to a cloud provider or a blockchain platform. In addition to deployment, design and implementation of blockchain-based applications is a hard task requiring deep expertise. Therefore, this paper presents a unified blockchain as a service platform (uBaaS) to support both design and deployment of blockchain-based applications. The services in uBaaS include \emph{deployment as a service}, \emph{design pattern as a service} and \emph{auxiliary services}. In uBaaS, \emph{deployment as a service} is platform agnostic, which can avoid lock-in to specific cloud platforms, while \emph{design pattern as a service} applies design patterns for data management and smart contract design to address the scalability and security issues of blockchain. The proposed solutions are evaluated using a real-world quality tracing use case in terms of feasibility and scalability. 
\end{abstract}


\end{frontmatter}


\section{Introduction}
\label{introduction}

Blockchain technology has attracted a wide range of interests as a distributed ledger technology for establishing digital trust in business. Many start-ups, enterprises, and governments \cite{ukreport,aureport} are currently exploring how to leverage blockchain technology to achieve trust and decentralization in the next generation of applications. Blockchain-based application areas are diverse, including physical or digital asset ownership management, tokens, currency, identity management, supply chain, electronic health records, voting, energy supply, and more.

Blockchain as a service (BaaS) is a promising solution to improve the productivity of blockchain application development. Easier deployment is the primary service offered by the existing BaaS platform providers, e.g. Microsoft Azure\footnote{\url{https://azure.microsoft.com/en-gb/solutions/blockchain/}\label{footnote:4}}, IBM\footnote{\url{https://www.ibm.com/blockchain/platform/}\label{footnote:5}}, and Amazon\footnote{\url{https://aws.amazon.com/blockchain/}\label{footnote:6}}. However, the existing BaaS deployment solutions are usually vendor locked, which are bound to either a cloud vendor (e.g. Microsoft Azure\textsuperscript{\ref{footnote:4}}) or a blockchain platform (e.g. IBM Hyperledger\textsuperscript{\ref{footnote:5}}). 

In addition to deployment, the design and implementation of blockchain-based applications are challenging to developers. First, blockchain is a new technology with limited tooling and documentation, so there can be a steep learning curve for developers. According to a survey by Gartner~\cite{Gartner:2018:CIOSurveyBC}, ``23 percent of [relevant surveyed] CIOs said that blockchain requires the most new skills to implement any technology area, while 18 percent said that blockchain skills are the most difficult to find.'' Second, blockchain is by design an immutable data store, so updating deployed blockchain smart contracts can be hard. This makes it difficult to fix bugs by releasing new versions of smart contracts. Mistakes in smart contracts have led to massive economic loss such as the DAO exploit on the Ethereum blockchain \cite{DAO}.\par
\
A software design pattern is defined as a solution to a problem that commonly occurs within a given context during software design \cite{Beck1987}. Design patterns for blockchain-based applications are best practices from industry and can be encapsulated services to ease the burden of developers and improve the quality of blockchain-based applications. One motivating example is the on-chain and off-chain design pattern, which provides a solution for storing data of blockchain-based applications. The purpose of separately storing data on-chain and off-chain is to ensure the integrity of on-chain data and privacy of the off-chain data. This pattern can be implemented as a service to generate an on-chain data registry smart contract and an off-chain data table in the conventional database based on the data model built by the developers. For example, in a quality tracing system, the on-chain attributes could be the traceability identifier and traceability result, while the off-chain attributes could be product price, buyer name, etc.

This paper presents a unified blockchain as a service platform named uBaaS, which facilitates the design and deployment of blockchain-based applications. The contributions of this paper are as follows.
\begin{itemize}[noitemsep,topsep=0pt]
	\item A set of design patterns for data management and smart contract design of blockchain-based applications to better take advantage of blockchain technology in practice. The design patterns include \emph{on-chain and off-chain}, \emph{hash integrity}, \emph{data encryption}, \emph{multiple authorities}, \emph{dynamic binding}, and \emph{embedded permission}.
    \item The uBaaS platform approach: 
    \begin{itemize}[noitemsep,topsep=0pt]
        \item \emph{Deployment as a service} which includes a blockchain deployment service and a smart contract deployment service. Deployment as a service can ease blockchain network deployment and  smart contract deployment. The proposed blockchain deployment service is platform agnostic, which can avoid lock-in to specific cloud platforms.
        \item \emph{Design pattern as a service} which consists of \emph{data management services} and \emph{smart contract design services}. Each service is designed based on a design pattern to better leverage the unique properties of blockchain (i.e. immutability and data integrity, transparency) and address the limitations (i.e. privacy and scalability).
	\end{itemize}
	\item Feasibility and scalability of the proposed solutions are evaluated using a real-world quality tracing use case. The evaluation results show that our solutions are feasible and have good scalability.
\end{itemize}
	
The remainder of this paper is organized as follows. Section 2 discusses the background and related work. Section 3 summarizes the design patterns for data management and smart contract design. Section 4 presents the overall architecture of uBaaS platform and design of each service provided by uBaaS. Section 5 introduces the implementation details. Section 6 evaluates the proposed solutions in terms of feasibility and scalability. Section 6 concludes the paper and outlines the future work.



\section{Related Work}

This section introduces the related work of our research. Blockchain technology and its classification are presented first, as it is the basis of this research. Further, how to apply blockchain technology to software systems is demonstrated, after which there is a brief introduction of the current obstacles to develop blockchain-based applications. Lastly design patterns and blockchain as a service are discussed.

\subsection{Blockchain Technology}
Blockchain is the technology behind Bitcoin \cite{Satoshi:bitcoin}, which is a decentralized data store that maintains all historical transactions of the Bitcoin network. The concepts of blockchain have been generalized to distributed ledger systems that verify and store transactions without coins or tokens \cite{scheuermann2015iacr}, without relying on any central trusted authority, e.g. traditional banking systems. Instead, all participants in the blockchain network can reach agreement on the states of transactional data to achieve trust.

The data structure of blockchain is an ordered list of identifiable blocks, each of which is connected to the previous block in the chain. Blocks are containers aggregating transactions while transactions are identifiable data packages that store parameters (such as monetary value) and function calls to smart contracts. A smart contract is a user-defined program that is deployed and executed on the blockchain network \cite{Omohundro:2014}, which can express triggers, conditions and business logic \cite{Weber:BPM2016} to enable more complex programmable transactions. Smart contracts can be implemented as part of transactions, and are executed across the blockchain network by all connected nodes. The blockchain platform Ethereum provides a built-in Turing-complete scripting language for writing smart contracts, called Solidity. The Ethereum Virtual Machine (EVM) is the execution environment for Ethereum, which comprises all full nodes on the network and executes bytecode compiled from Solidity scripts. 

\begin{table*}[t]
\centering
\caption{Types of Blockchain ($\onePlusses$: Less favourable, $\twoPlusses$: Neutral, $\threePlusses$: More favourable)\\}
\label{tab:blockchain}
\resizebox{\textwidth}{21mm}{
\begin{tabular}{p{0.31\columnwidth}p{0.17\columnwidth}p{0.17\columnwidth}p{0.17\columnwidth}p{0.17\columnwidth}}
\toprule

\multicolumn{1}{c}{\multirow{3}{0.31\columnwidth}{\bf \centering Type}} & \multicolumn{4}{c}{\bf \centering Impact} \\
\cmidrule(l){2-5}
& \multirow{2}{0.17\columnwidth}{\centering \bf Fundamental properties} & \multirow{2}{0.17\columnwidth}{\centering \bf Cost efficiency} &  \multirow{2}{0.17\columnwidth}{\centering \bf Performance} &  \multirow{2}{0.17\columnwidth}{\centering \bf Flexibility}\\ \\
\midrule

Public blockchain & \multicolumn{1}{c}{$\threePlusses$} & \multicolumn{1}{c}{$\onePlusses$} & \multicolumn{1}{c}{$\onePlusses$} & \multicolumn{1}{c}{$\onePlusses$}\\ \cmidrule(l){1-5} 
Consortium blockchain & \multicolumn{1}{c}{$\twoPlusses$} & \multicolumn{1}{c}{$\twoPlusses$} & \multicolumn{1}{c}{$\twoPlusses$} & \multicolumn{1}{c}{$\twoPlusses$}\\ \cmidrule(l){1-5} 
Private blockchain & \multicolumn{1}{c}{$\onePlusses$} & \multicolumn{1}{c}{$\threePlusses$} & \multicolumn{1}{c}{$\threePlusses$} & \multicolumn{1}{c}{$\threePlusses$}\\ 
\bottomrule
\end{tabular}}
\end{table*}

\subsection{Types of Blockchain}

As shown in Table \ref{tab:blockchain}, there are three types of blockchain in terms of deployment, including public blockchain, consortium blockchain, and private blockchain. Public blockchains, which are used by most digital currencies, can be accessed by anyone on the Internet. Using a public blockchain achieves better data transparency and auditability, but sacrifices performance and has a different cost model. A consortium blockchain is typically used across multiple organizations and has pre-authorized nodes to control the consensus process. In a private blockchain network, write permissions are often kept within one organization, although this may include multiple divisions of a single organization. The right to read the consortium or private blockchain may be public or may be restricted to a specific group of participants. When using a consortium or private blockchain, a permission management component is required to authorize participants within the blockchain network. Private blockchains are the most flexible for configuration because the network is governed and hosted by a single organization. Our work is mainly focused on consortium blockchains and private blockchains.

\subsection{Blockchain as a Component of Software Application Systems}
Researchers in academia and developers in industry are investigating and exploring how to build next generation applications using blockchain technology \cite{ukreport,aureport}. Application areas in industry include but are not limited to digital currency, international payments, registries, government identity and taxation management, Internet of Things (IoT) identify and security management, and supply chain \cite{LI2017,REYNA2018173,lu2017adaptable, XU2019399}. Furthermore, there is various academic work that exploits blockchain to address issues in different domains. Qu et al.~\cite{QU2019208} propose spatio-temporal blockchain technology that supports fast query processing, which is proved to be applicable and effective. Zyskind et al. use blockchain to build a personal data management system that ensures users own and control their data in a decentralized way\cite{decentralizePrivacy}. Bulat Nasrulin et al. proposed a mobility analytics application that is built on top of blockchain\cite{DBLP:conf/mdm/NasrulinMQ18} and renovated blockchain with distributed database\cite{MMuzammal2018}. Namecoin \cite{blockstack2016} and PKI \cite{IKP} are two public key platforms built on blockchain. ProvChain enables data Provenance for cloud-based data analytics\cite{ProveChain}. 

\begin{figure}[t]
\begin{center}
\centerline{\includegraphics{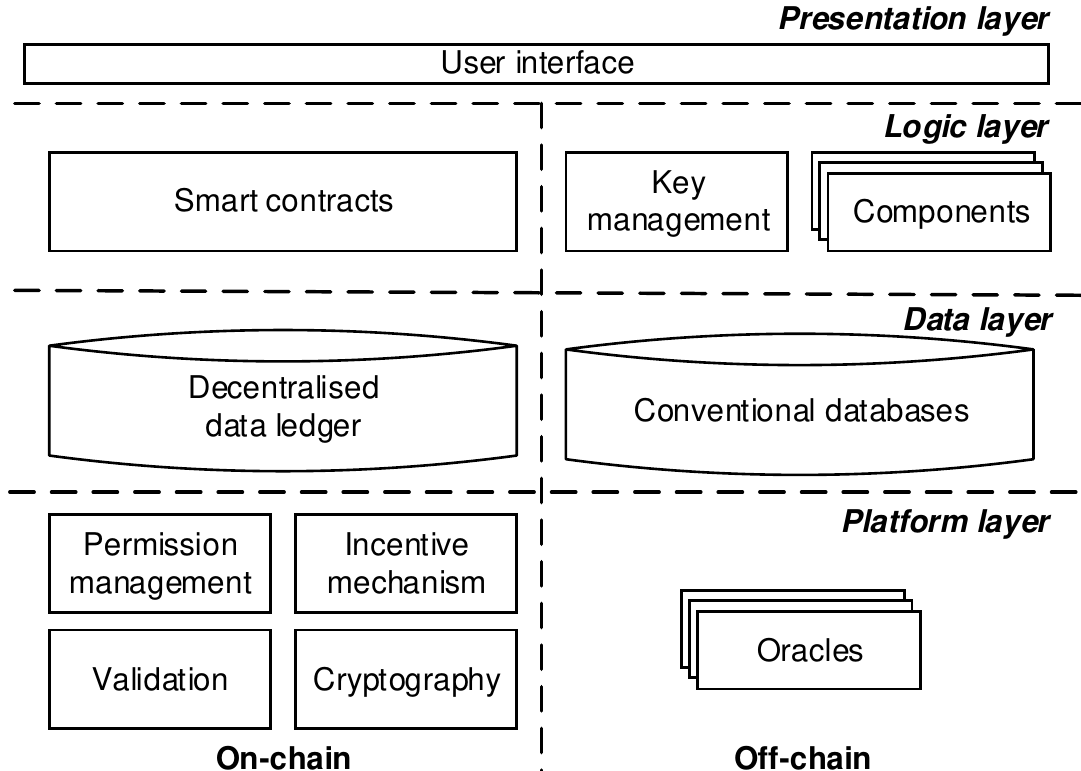}}
\caption{Blockchain as a component of software application systems.}
\label{blockchainapplication}
\end{center}
\end{figure}

Software components are the fundamental building blocks for software architecture,
and blockchain can be a software component offering computational capabilities. Blockchains are complex, network-based software components, which can provide data storage, computation services, and communication services.

Fig.~\ref{blockchainapplication} illustrates the architecture of a blockchain-based application system, which consists of four horizontal layers and two vertical layers. The horizontal layers include a presentation layer, a logic layer, a data layer, and a platform layer, while the vertical layers are on-chain and off-chain.

The users interact with blockchain smart contracts and off-chain components (including the key management component) via the user interface. The blockchain-based application system can implement business logic through different off-chain components and on-chain smart contracts. Key management is an essential off-chain component in such blockchain-based system. Every participant in a blockchain network has one or more private keys, which are used to digitally sign the transactions. The security of these private keys is important: if the private key is stolen, assets held by the respective account can be accessed and protected functions of smart contracts can be invoked.

At the data layer, blockchain acts as a decentralized data ledger to store data which requires integrity and/or transparency. Off-chain conventional databases are often needed to store private or large-size data due to the scalability and privacy.

At the platform layer, the fundamental blockchain features can include permission management, incentive mechanisms, transaction validation, and cryptographically-secure payment. The oracles supply information about the external world to the blockchain, usually by adding that information to the blockchain as data in a transaction.

\subsection{Challenges of Blockchain Application Development}
Blockchain is an emerging distributed ledger technology which requires developers to learn new skills and have a deep understanding of the technology in order to build blockchain-based applications. We summarize the challenges of blockchain application development as follows.

\begin{itemize}[noitemsep,topsep=0pt]
	\item Deployment. The current blockchain deployment solutions (e.g. Microsoft Azure\textsuperscript{\ref{footnote:4}}, IBM Hyperledger\textsuperscript{\ref{footnote:5}}, Amazon\textsuperscript{\ref{footnote:6}}) lock customers in to specific cloud and/or blockchain platforms. However, many enterprises or governments require building blockchain-based applications on their own on-premise private cloud, which are not met by the existing blockchain solutions. Besides, the deployment process of blockchain is error-prone, time-consuming, and requires frequent updating.
	\item Scalability. Blockchain has limited storage capability since it contains a full history of all the transactions across all participants of the blockchain network. Thus, the size of blockchain continues to grow. The ever-growing size of blockchain is a challenge for storing data on blockchain. Also, storing large amounts of data or deploying large smart contracts within a transaction may be impossible due to the limited size of the blocks of the blockchain, which is under control of the network~\cite{SRDS2017}. 
    \item Data privacy. Blockchain-based applications might have sensitive data which should be only available to some certain blockchain participants. However, the information on blockchain is designed to be accessible to all the participants. There is no privileged user within the blockchain network, whether the blockchain is public, consortium or private. 
    \item Key management. Authentication on blockchain is achieved by digital signatures. However, blockchain does not offer any mechanism to recover a lost or a compromised private key. Losing a key results in permanent loss of control over an account, and potentially smart contracts that refer to it.
    \item Permission control. All the smart contracts deployed on blockchain can be accessed and called by all the blockchain participants by default. A permission-less function might be triggered by unauthorized users accidentally, which becomes a vulnerability of blockchain-based application.  
\end{itemize}

 \subsection{Design Patterns and Blockchain as a Service}
A design pattern is a reusable solution to a problem that commonly occurs within a given context during software design \cite{gof}. There are a few works on design patterns for blockchain-based applications. J. Eberhardt and S. Tai present a group of patterns which mainly focus on on-chain and off-chain data and computation \cite{on-off-chain}. Zhang et al. applies four existing object-oriented software patterns to smart contract design in the context of a blockchain-based health care application \cite{factorypattern}. Liu et al. summarized eight smart contract design patterns and classified them into four categories\cite{liu2018applying}. Bartoletti and Pompianu conduct an empirical analysis of smart contracts, in which they collected hundreds of smart contracts and divided them into several categories: token, authorization, oracle, randomness, poll, time constraint, termination, math and fork check \cite{empiricalAnalysis}. 
In recent work, some of the authors of this paper proposed an extensive pattern collection~\cite{EuroPLoP-2018}.
However, the summarized design patterns for blockchain-based applications are mostly at conceptual level, requiring users to implement the solutions.

Blockchain as a service (BaaS) provides an offering that allows developers to develop blockchain-based applications efficiently.  Most of the current BaaS platforms are designed to help developers create, deploy, and manage blockchain network, e.g. 
Microsoft Azure\textsuperscript{\ref{footnote:4}} , IBM\textsuperscript{\ref{footnote:5}}, and Amazon\textsuperscript{\ref{footnote:6}}. 
However, the existing BaaS deployment solutions are usually locked into a specific public cloud or blockchain network provider (e.g. Microsoft Azure\textsuperscript{\ref{footnote:4}}, IBM Hyperledger\textsuperscript{\ref{footnote:5}}). Many governments or enterprises deploy their applications in private clouds and may have different blockchain network preferences. Besides, in addition to deployment services, development services are also required for BaaS platforms. Architecture design of blockchain application is a challenge to developers since they need to learn blockchain programming languages and have a deep understanding of blockchain technology.

\section{Design Patterns for Blockchain-based Applications}

In software engineering, a design pattern is a reusable solution to a problem that commonly occurs within a given context during software design \cite{Beck1987}. In this section, we summarize and categorize six design patterns which can be applied to the architecture design of a blockchain-based application system. Those patterns are divided into \emph{data management design patterns} and \emph{smart contract design patterns} according to their characteristics and effects.

\subsection{Data Management Design Patterns}
As discussed in Section 2.3, in a blockchain-based application system, blockchain can act as a decentralized data ledger and work with conventional databases to store data. Data management of blockchain-based application is challenging due to the fundamental properties (e.g. data transparency) and limitations of blockchain (e.g. blockchain scalability). Here we summarize three design patterns for data management: \emph{on-Chain and off-Chain}, \emph{hash integrity} and \emph{data encryption}.

\subsubsection{On-Chain and Off-Chain}
\vspace{0.5em}\noindent \textbf{Summary:} The on-chain and off-chain pattern separately stores data on blockchain and off blockchain to ensure the data integrity while addressing the storage capability issue of blockchain.

\vspace{0.5em}\noindent \textbf{Context:} Some applications consider leveraging blockchain to ensure data integrity since all the data on blockchain are transparent and immutable.

\vspace{0.5em}\noindent \textbf{Problem:} It may be impossible to store sensitive and large amounts of data on blockchain since all the information on blockchain is accessible to all participants and blockchain has limited storage capacity.
 
\vspace{0.5em}\noindent \textbf{Solution:} 
Rather than placing all the data on-chain, only the sensitive and small data is stored on-chain. For example, a food quality tracing system can store traceability information that is required by the traceability regulation (e.g. traceability number and results) on-chain, while places the factory production process photos off-chain. The benefits of separately storing data on blockchain and off blockchain is to better leverage the properties of blockchain and avoid the limitations of blockchain. Blockchain can guarantee the integrity and immutability of the critical data on-chain. The non-tiny files are stored off-chain so that the size of the blockchain would not grow so fast. Storing the hashes of off-chain files can further ensure the integrity of the off-chain files.

\subsubsection{Hash Integrity}
\vspace{0.5em}\noindent \textbf{Summary:} The hash integrity pattern uses hashing to ensure the integrity of arbitrarily large datasets which may not fit directly on the blockchain.

\vspace{0.5em}\noindent \textbf{Context:} Some blockchain-based applications consider using blockchain to ensure the integrity of large amounts of data.

\vspace{0.5em}\noindent \textbf{Problem:} It may be impossible to store large amounts of data within a transaction since the blocks of blockchain has limited size (e.g., Ethereum has a block gas limit to control the data size, computational complexity and number of transactions included in a block). There is a problem about how to store arbitrary size data on blockchain to guarantee data integrity. 
\vspace{0.5em}\noindent \textbf{Solution:} For data of large size (essentially data that is bigger than its hash value), rather than storing the raw data directly on blockchain, a hash value of the raw data is stored on blockchain. The hash value is produced by a hash function which maps data of arbitrary size to data of fixed size and is non-invertible. Any change to the data will lead to a change in its corresponding hash value.

\subsubsection{Data Encryption}
\vspace{0.5em}\noindent \textbf{Summary:} The data encryption pattern ensures confidentiality of the data stored on blockchain by encrypting it.

\vspace{0.5em}\noindent \textbf{Context:} 
For some blockchain-based applications, commercially sensitive data should be only accessed by specific participants. An example would be a special discount price offered by a service provider to a subset of its users. Such information might not be supposed to be accessible to the other users who do not get the discount.

\vspace{0.5em}\noindent \textbf{Problem:} 
The lack of data privacy is one of the main limitations of blockchain. All the information on blockchain is publicly available to the participants of the blockchain. There is no privileged user within the blockchain network, no matter the blockchain is public, consortium or private. On a public blockchain, new participants can join the blockchain network freely and access all the information recorded on blockchain. Any confidential data on public blockchain is exposed to the public.

\vspace{0.5em}\noindent \textbf{Solution:} 
Asymmetric(or symmetric) encryption can be used to encrypt data before storing the data on blockchain. One of the involved participants generate a key pair and distribute the decryption key to other involved participants. The involved participants can encrypt the data before placing it on blockchain using the encryption key. Only the involved participants who have the decryption key can decrypt the data.

\subsection{Smart Contract Design Patterns}
Developers can define smart contracts (i.e. software programs on blockchain) to implement business logic and enable more complex programmable transactions. However, developing smart contracts usually require developers to have a deep understanding of blockchain and rich smart contract development experience. Thus, we summarize three design patterns as solutions to address the common problems (e.g. security) in the design of smart contracts.

\subsubsection{Multiple Authorities}
\vspace{0.5em}\noindent \textbf{Summary:} 
A set of blockchain account addresses which can authorize a transaction is pre-defined. Only a subset of the pre-defined account addresses is required to authorize transactions.\par

\vspace{0.5em}\noindent \textbf{Context:} 
Some activities in blockchain-based applications might need to be authorized by multiple parties (i.e. blockchain account addresses). For example, a monetary transaction may require authorization from multiple blockchain account addresses.

\vspace{0.5em}\noindent \textbf{Problem:} 
The actual addresses that authorize an activity might not be able to be determined due to the availability of the authorities.

\vspace{0.5em}\noindent \textbf{Solution:} 
To enable more flexible binding, an M-of-N mechanism can be used to define that M out of N private keys are required to authorize the transaction. M is the threshold of authorization. 

\subsubsection{Dynamic Binding}

\vspace{0.5em}\noindent \textbf{Summary:} 
The dynamic binding pattern uses a hash created off-chain to dynamically bind authority for a transaction.\par
\vspace{0.5em}\noindent \textbf{Context:} 
In blockchain-based applications, some activities need to be authorized by one or more participants that are unknown when the corresponding smart contract is deployed or the transaction is submitted to blockchain. 

\vspace{0.5em}\noindent \textbf{Problem:} 
Blockchain does not support dynamic binding with a blockchain account address which is not defined in the transaction or smart contract. All accounts that can authorize a second transaction have to be defined in the first transaction before that transaction is added to the blockchain. 

\vspace{0.5em}\noindent \textbf{Solution:} 
An off-chain secret can be used to enable a dynamic binding when the participant authorizing a transaction is unknown beforehand. In the context of payment, when the sender deposits money to an escrow smart contract, a hash of a secret is submitted with the money as well. The participant who receives the secret off-chain can claim the money from the escrow smart contract by revealing the secret. Thus, the receiver of the money does not need to be defined beforehand in the escrow contract. 

\subsubsection{Embedded Permission}

\vspace{0.5em}\noindent \textbf{Summary:} 
Smart contracts use an embedded permission control to restrict access to the invocation of the functions defined in the smart contracts.\par
\vspace{0.5em}\noindent \textbf{Context:} 
A smart contract by default has no owner, since the smart contracts running on blockchain can be accessed and called by all the blockchain participants and other smart contracts by default.

\vspace{0.5em}\noindent \textbf{Problem:} 
Once the smart contract is deployed, the author of the smart contract has no special privilege to invoke on the smart contract. A permission-less function can be triggered by unauthorized users accidentally, which becomes a vulnerability of blockchain-based application. For example, a permission-less function is discovered in a smart contract library used by the Parity multi-sig wallet, caused the freezing of about 500K Ether\footnote{\url{https://paritytech.io/a-postmortem-on-the-parity-multi-sig-library-self-destruct/}}. In 2016, 7\% smart contract on public Ethereum could be terminated without authority~\cite{NewKids-2016}.

\vspace{0.5em}\noindent \textbf{Solution:} 
Permission control can be added to every smart contract function to check permissions for every function caller based on the blockchain address of the caller before executing the logic of the function. Calls from unauthorized blockchain addresses are rejected.

\begin{figure}[h!]
\begin{center}
\centerline{\includegraphics[width = 1.0\textwidth]{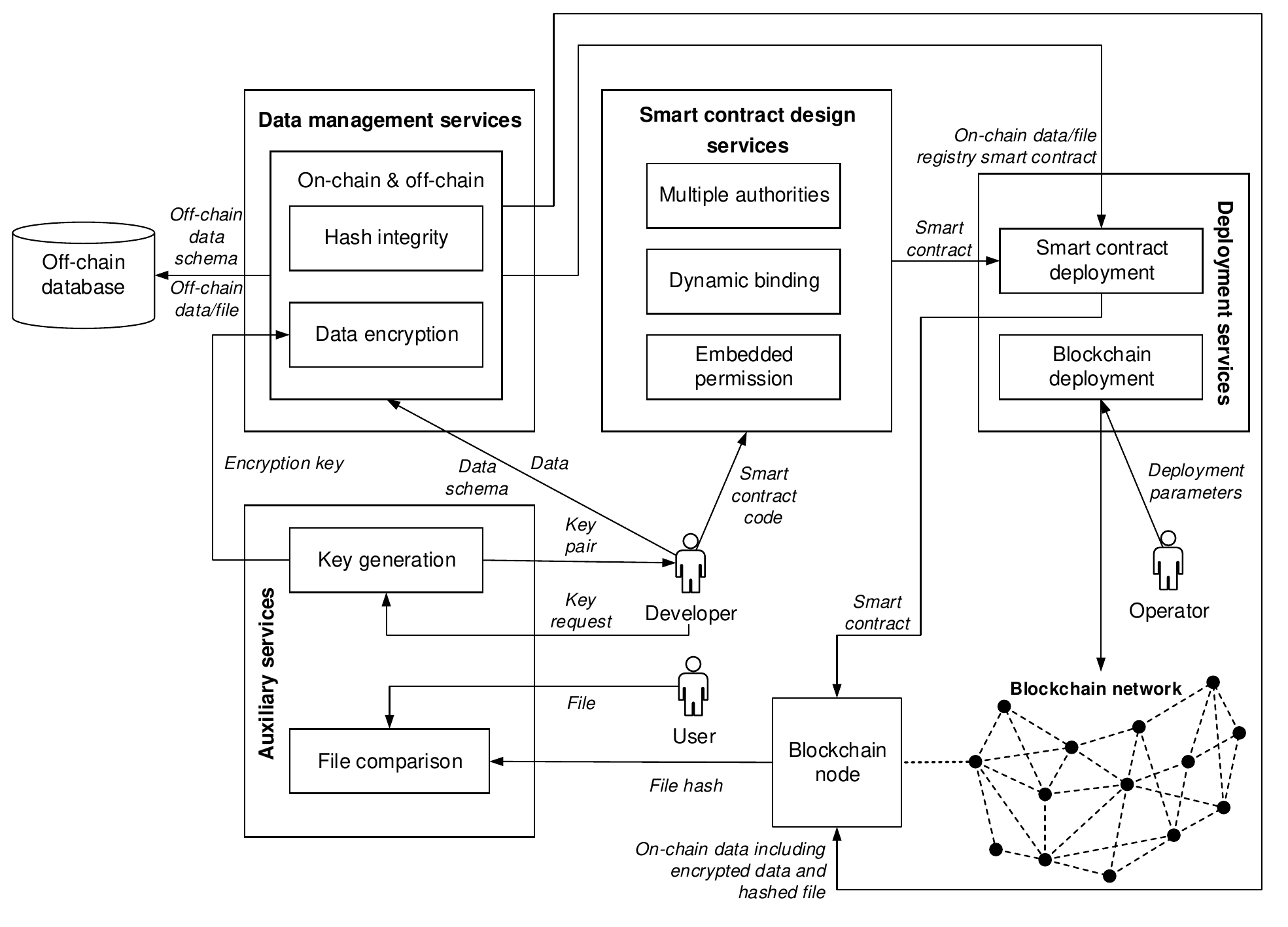}}
\caption{Architecture of uBaaS.}
\label{architecture}
\end{center}
\end{figure}

\section{Architecture of uBaaS}
In uBaaS, we propose deployment as a service for vendor-independent deployment and design pattern as a service to address the scalability and security issues of blockchain-based applications. Fig.~\ref{architecture} illustrates the overall architecture of uBaaS. The services proposed in uBaaS are classified into three categories, \emph{deployment as a service}, \emph{design pattern as a service} and \emph{auxiliary services}. Users can build up blockchain environment and design blockchain-based applications via uBaaS front-end user interface, which interacts with the back-end services through an API gateway. The API gateway forwards API calls from the front-end user interface to the corresponding services.

The remainder of this section introduces \emph{deployment as a service}, \emph{design pattern as a service} and \emph{auxiliary services} respectively. \emph{Deployment as a Service} consists of 2 kinds of deployment services, \emph{design pattern as a service} includes 6 design pattern services to help developers take advantage of blockchain's properties and address blockchain's limitations, while \emph{auxiliary services} can act as assistance to better leverage design pattern as a service.

\subsection{Deployment as a Service}
\emph{Deployment as a service} includes \emph{blockchain deployment service} and \emph{smart contract deployment service}. Users can configure the blockchain settings (such as difficulty and participant node IP) and deploy their customized blockchain network using uBaaS. Infrastructure-as-code is applied to the blockchain deployment service which enables the automation of deployment, configuration, and task management by the developed script. Once the blockchain is set up, users can monitor the status of blockchain at real-time and deploy the developed smart contracts on the blockchain. We assume we can access the participant nodes since we focus on consortium blockchain and private blockchain. Currently, both Ethereum blockchain and Hyperledger Fabric blockchain are supported. We plan to add more blockchain platforms as deployment options in uBaaS.

The smart contract deployment service enables users to select the developed smart contract (i.e. program) file and deploy the smart contract code on the blockchain. Besides, the smart contract address and Application Binary Interface (ABI) are stored in the off-chain database for invoking the deployed smart contract.

\subsection{Design Pattern as a Service}
\emph{Design pattern as a service} consists of \emph{data management services} and \emph{smart contract design services}. \emph{Data management services} include \emph{on-chain and off-chain service}, \emph{data encryption service}, and \emph{hash integrity service}, while \emph{smart contract design services} comprises \emph{multiple authorities service}, \emph{dynamic binding service}, \emph{embedded permission service}. Each type of services is designed based on a design pattern to improve scalability, adaptability, and security of blockchain-based applications. The \emph{data management services} enable users to manage data directly via uBaaS front-end user interface while the \emph{smart contract design services} integrate smart contract design patterns into the original smart contract. 

\subsubsection{Data Management Services}~\\
To address the issues of blockchain storage capability limitation and data privacy, an \emph{on-chain and off-chain service} is proposed to store the critical data which is required to be immutable on-chain while keep all the data off-chain to enhance the data reading efficiency. Users can define the data schema and determine which attributes are stored on-chain and off-chain. Based on the data model built by users, the service builds up a data registry on blockchain by deploying the generated on-chain data registry smart contract and sets up an off-chain data table in the conventional database. The on-chain data registry and off-chain data table share the same name. The user can write/read data to/from the selected data store regardless to being stored on-chain or off-chain.  

To preserve the privacy of the involved participants, uBaaS provides a \emph{data encryption service} that encrypts on-chain data to ensure confidentiality of the data stored on blockchain. The uBaaS user first encrypts the data item using the private key (generated by the key generation service in auxiliary services) and then stores it on blockchain. The blockchain participants who have the public key are allowed to access the transaction and decrypt the information. By using the on-chain data encryption service, the sensitive data stored on blockchain are not accessible to blockchain participants who do not hold the public key.

To store the large size data (i.e. file) on blockchain, the \emph{hash integrity service} generates the hash value of the file and stores the hash in the on-chain file registry which is associated with the on-chain data registry using the pre-defined foreign key.

\subsubsection{Smart Contract Design Services}~\\
The \emph{smart contract design services} focus on the permission control for invocation of the smart contract functions. The input of each of those services is the smart contract code written by the user while the output is updated smart contract code by adding the template code implementing the corresponding smart contract design pattern.

The \emph{multiple authorities service} focuses on the smart contact functions that can be invoked only when the authorized blockchain addresses approve. Users can predefine a group of blockchain addresses which can authorize a transaction (i.e. calling a function in the smart contract) and set the minimal number of authorizations for transaction approval. The users need to select a smart contract they write and the function which needs the mechanism of multiple authorities. Then uBaaS modifies the code of the selected function and generate an updated smart contract with code for multiple authorities. 

The \emph{dynamic binding service} uses an off-chain secret to enable a dynamic authorization when the participant approving a transaction (i.e. calling a function in the smart contract) is unknown beforehand. Users need to provide an secret (e.g. a random number) and select the corresponding function in the smart contract. The service modifies the function code by adding the hash of the secret and generate the updated smart contract. There is no need for a special protocol to exchange the secret as it can be exchanged in any ways off-chain. Only the user who has the secret can invoke the selected function in the smart contract.

The purpose of \emph{embedded permission service} is to restrict access to the invocation of the functions defined in the smart contracts. Users can identify the authorities for the selected function in the smart contract by providing the authority addresses. The service adds permission control code to the smart contract function to check permissions for every caller based on the blockchain addresses of the caller, which is done before executing the function logic. 

\begin{figure}[htbp]
\begin{center}
\centerline{\includegraphics[width = \textwidth]{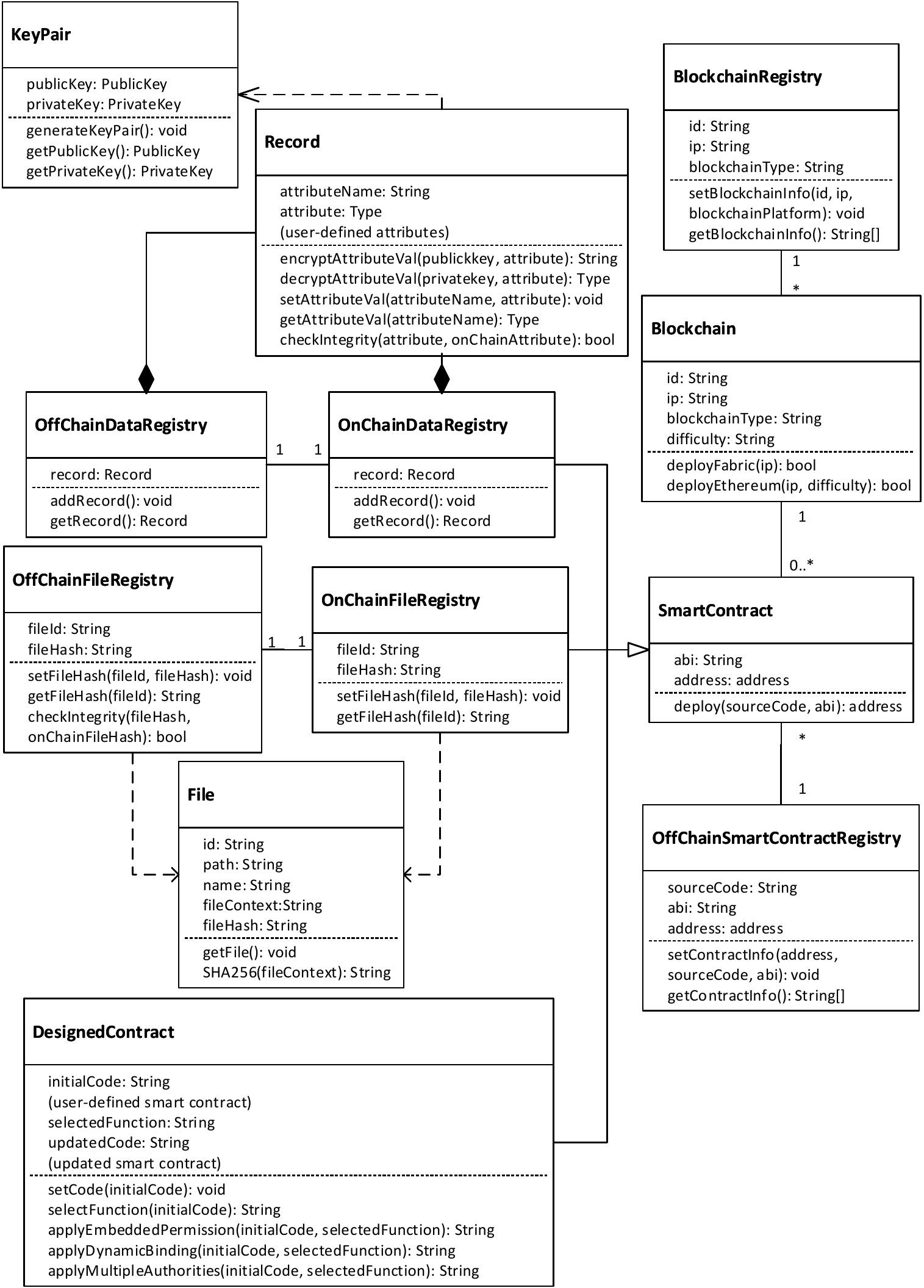}}
\caption{The main class diagram of uBaaS implementation.}
\label{classDiagram}
\end{center}
\end{figure}

\subsection{Auxiliary Services}
The current version of uBaaS supports two auxiliary services, \emph{key management service} and \emph{file comparison service}. 

The \emph{key management service} is used to generate key pairs for data encryption. The developers can encrypt the data using the data encryption service and share the decryption key with other platform users (e.g. developers or application users) before store the encrypted data in the on-chain data registry. The channel for sharing the decryption key is out of the scope in this paper.  

The purpose of \emph{file comparison service} is to validate the authenticity of a file (e.g. higher education certificates). Users can select the file from local node and check the authenticity of the file by comparing its hash value with the hash value of the associated original file stored in the on-chain file registry.

\begin{figure}[t]
\begin{center}
\centerline{\includegraphics[width = \textwidth]{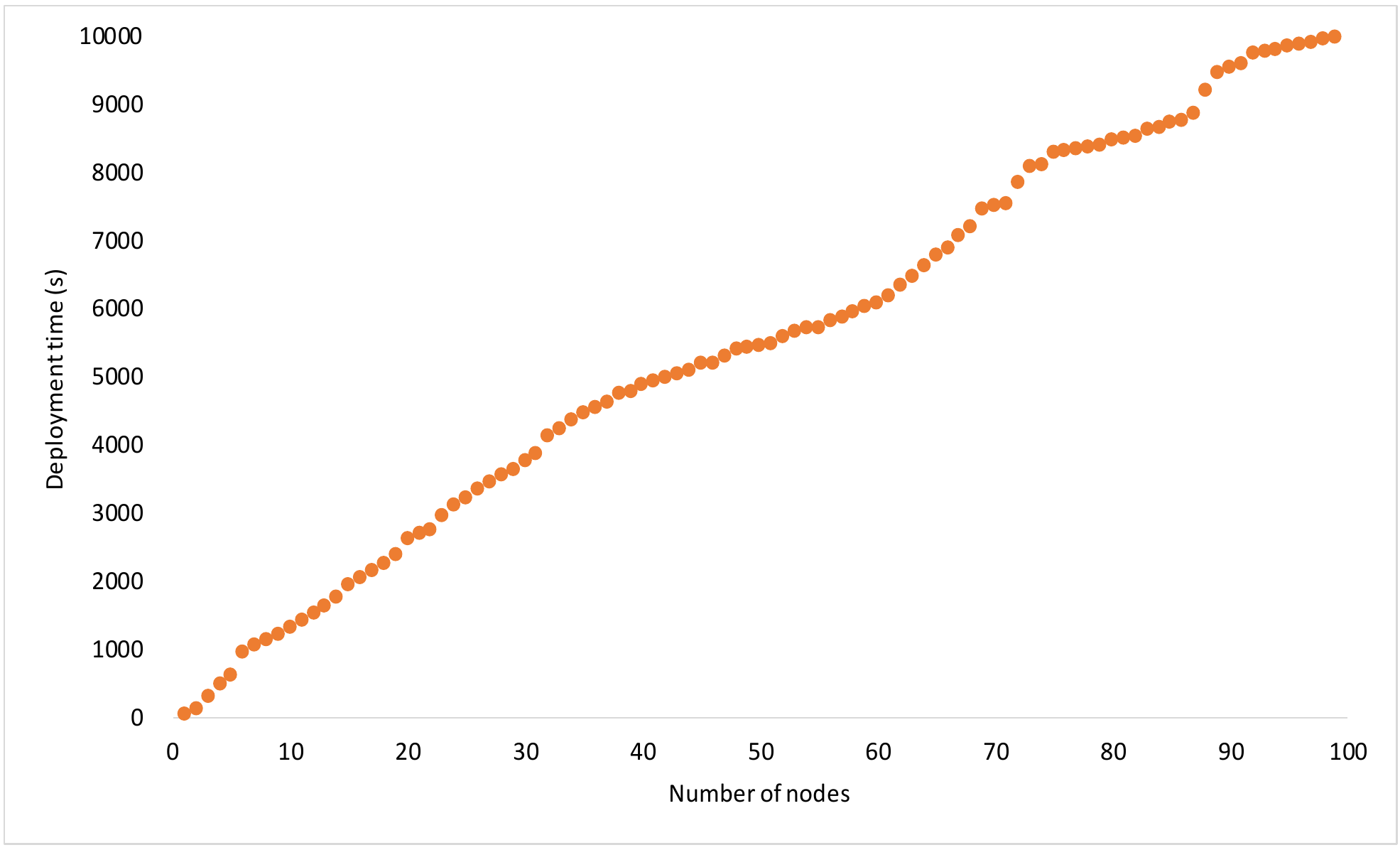}}
\caption{Blockchain deployment time.}
\label{eDeployment}
\end{center}
\end{figure}

\begin{figure}[t]
\begin{center}
\centerline{\includegraphics[width = \textwidth]{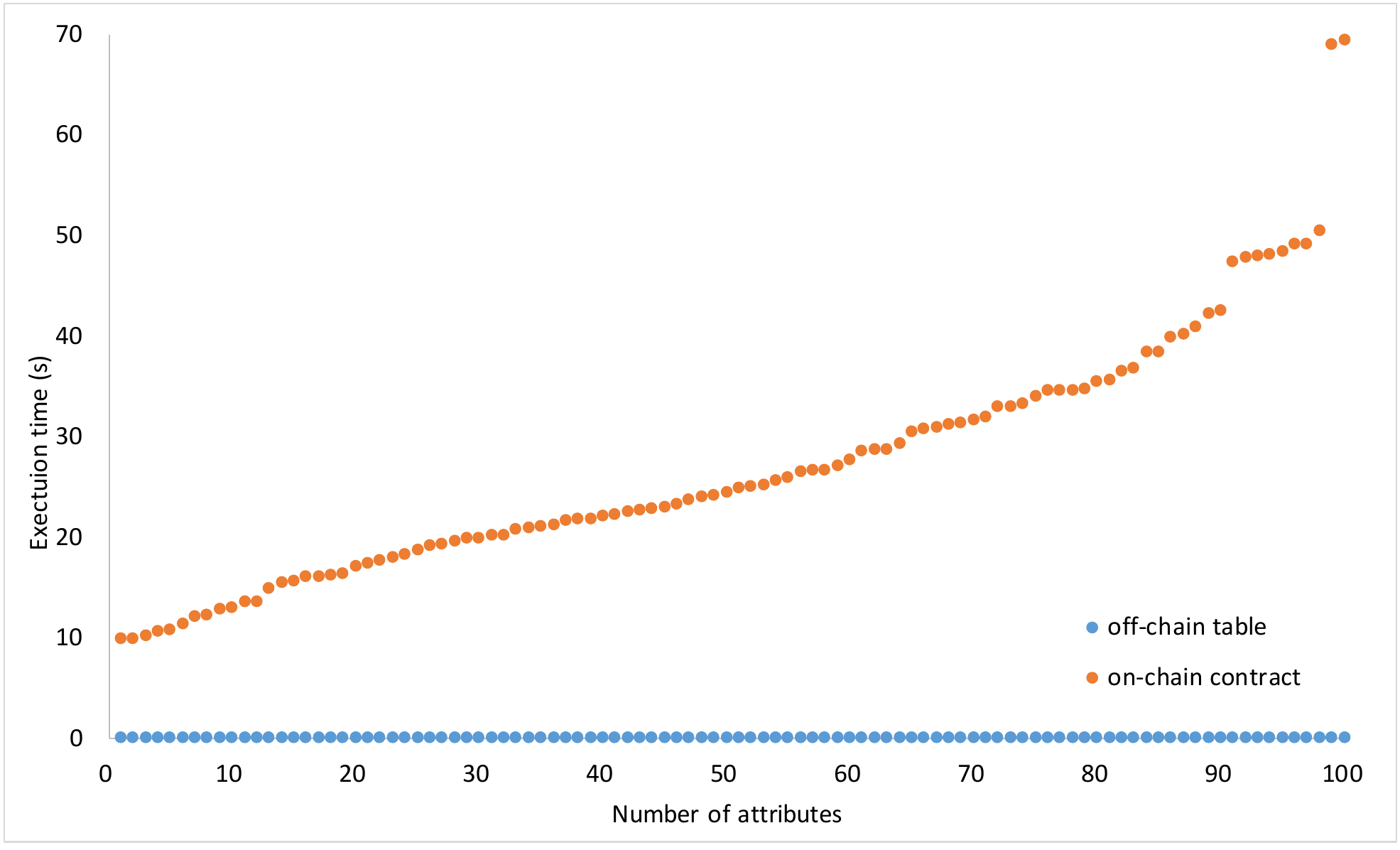}}
\caption{Execution time of generating an off-chain data table and an on-chain data registry smart contract.}
\label{createTable}
\end{center}
\end{figure}

\begin{figure}[t]
\begin{center}
\centerline{\includegraphics[width = \textwidth]{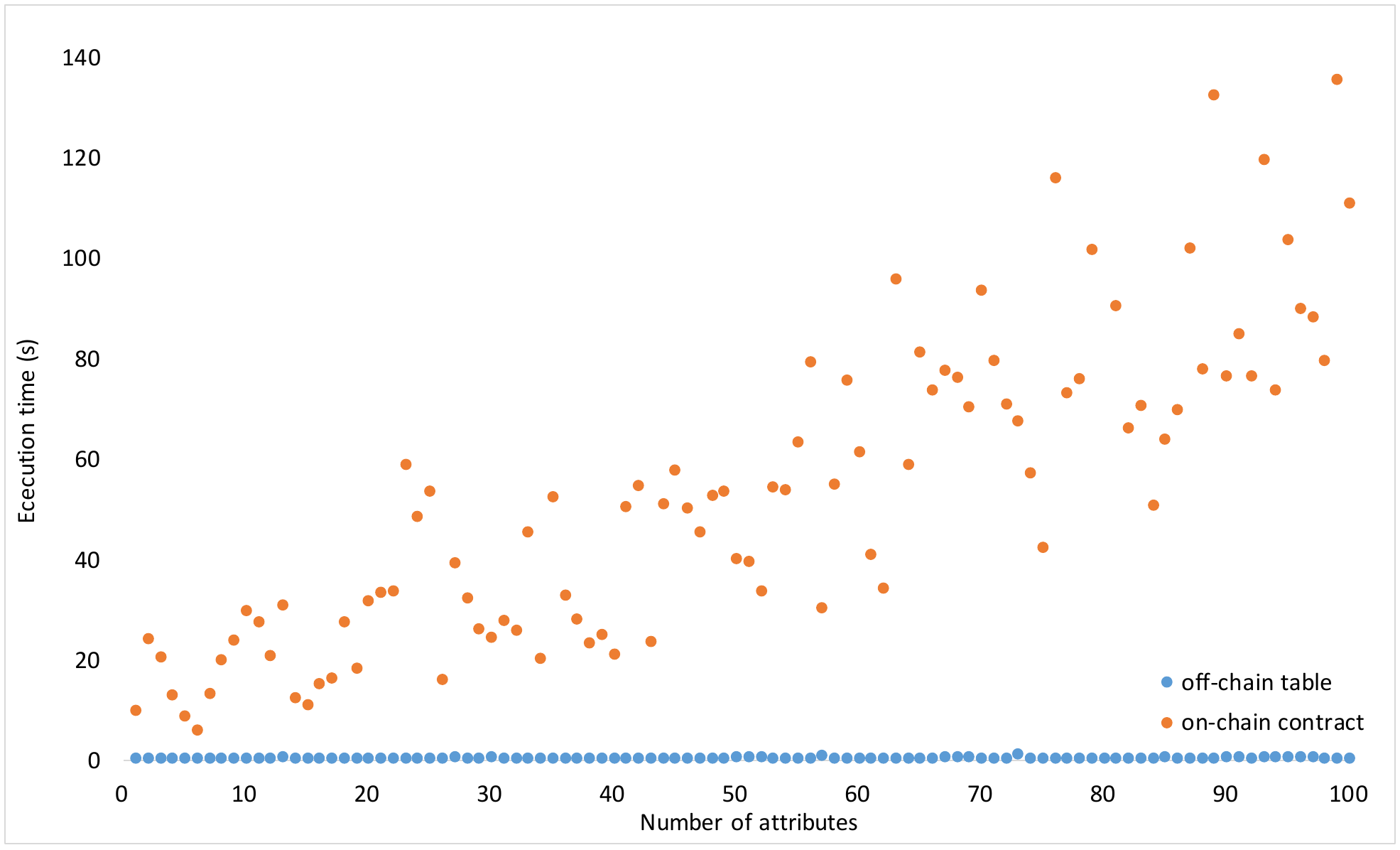}}
\caption{Execution time of writing data to the off-chain data table and the on-chain data registry smart contract.}
\label{writingData}
\end{center}
\end{figure}

\begin{figure}[t]
\begin{center}
\centerline{\includegraphics[width = \textwidth]{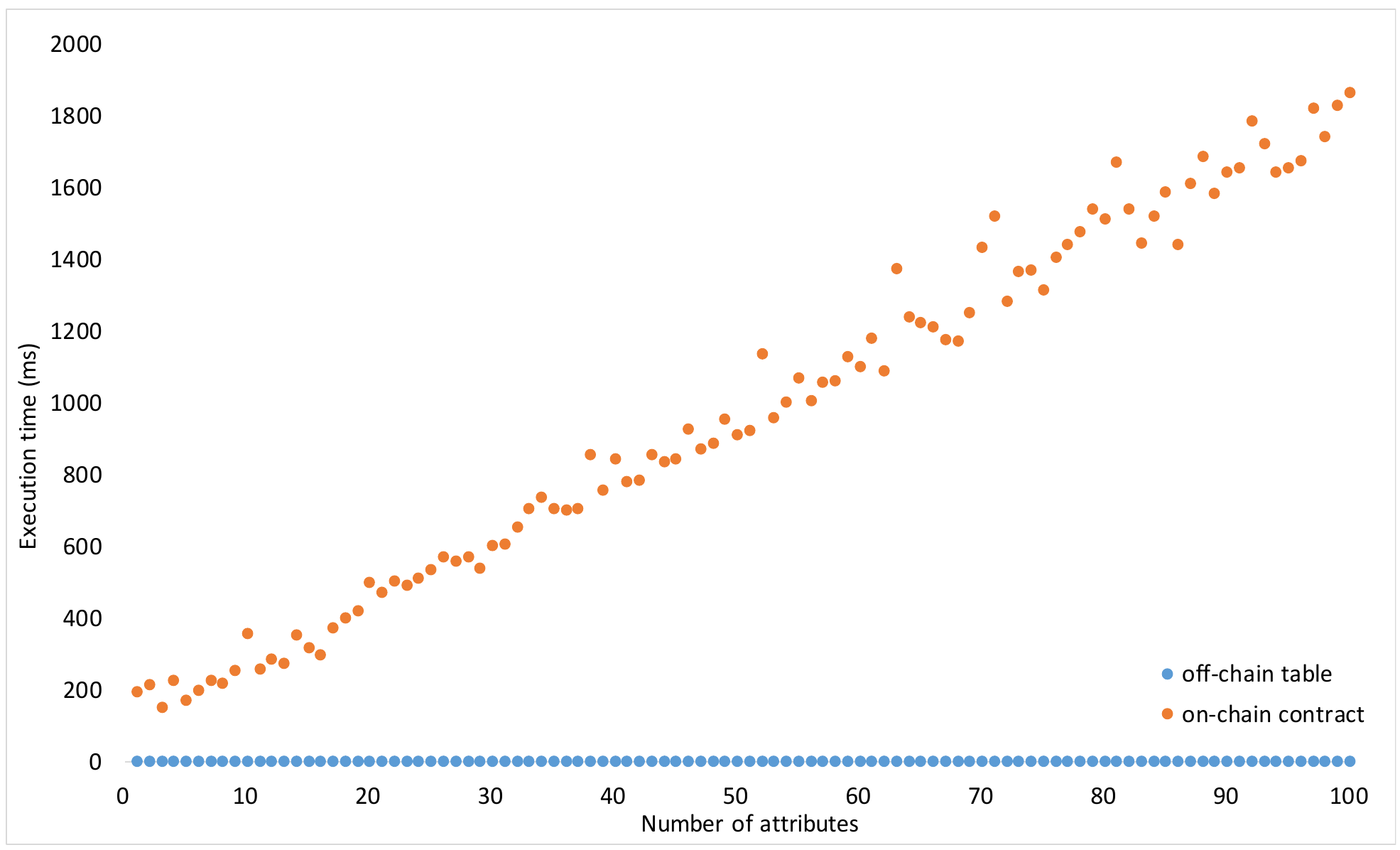}}
\caption{Execution time of reading data from the off-chain table and the on-chain smart contract.}
\label{queryingData}
\end{center}
\end{figure}

\begin{figure}[t]
\begin{center}
\centerline{\includegraphics[width = 0.9 \textwidth]{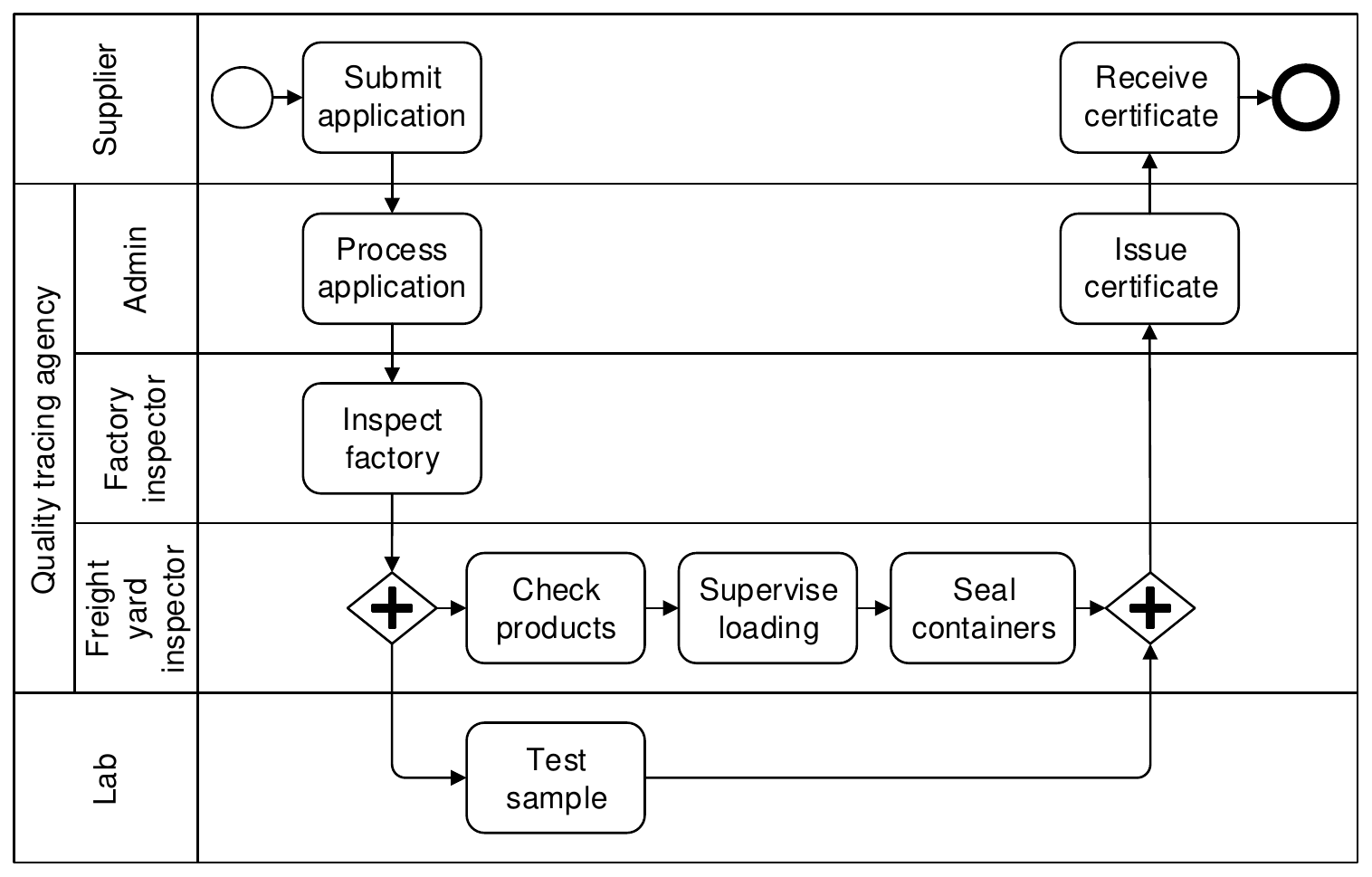}}
\caption{Quality tracing process.}
\label{qualitytracing}
\end{center}
\end{figure}

\section{Implementation}

Fig.~\ref{classDiagram} illustrates the implementation design of uBaaS using the class diagram. \emph{BlockchainRegistry} maintains each deployed \emph{Blockchain} network, while \emph{OffChainSmartContractRegistry} stores the source code and information of \emph{SmartContract}. There are three classes that inherit from \emph{SmartContract}: \emph{OnChainDataRegistry}, \emph{OnChainFileRegistry} and \emph{DesignedContract}. Each \emph{OnChainDataRegistry} is associated with an \emph{OffChainDataRegistry} for data storage. User-defined data record attributes are contained in \emph{Record}, and \emph{OnChainDataRegistry} and \emph{OffChainDataRegistry} are composed of \emph{Record}. Note that the attribute values stored in \emph{Record} must be encrypted using \emph{KeyPair} before storing in \emph{OnChainDataRegistry}. \emph{File} is stored in both \emph{OffChainFileRegistry} and its hash value is stored in \emph{OnChainFileRegistry}. \emph{DesignedContract} applies smart contract design patterns to the smart contract code written by users.

The platform is developed in Java 1.8 using Eclipse Java IDE 4.6.0 and released using Tomcat v7.0 server. To achieve blockchain deployment as a service, we used SSH to transfer the deployment files, including the client file (e.g. Geth for Ethereum) and genesis block file (e.g. genesis.json for Ethereum), to the participant nodes. To implement data management services, we selected MySQL 5.7.17 as the supported database to store off-chain data. Regarding smart contract design services, the smart contract design pattern templates are pre-developed and stored in the database. The platform users can select the smart contract design pattern that they want to apply. For both data management services and smart contract design services, the smart contracts are written in Solidity, compiled with Solidity compiler version 0.4.24.  After compilation, a smart contract is deployed on blockchain via web3 API, returning smart contract address as result if the deployment succeeds.

\section{Evaluation}
In this section, we evaluate the feasibility and scalability of uBaaS. We first introduce the experiment environment. Then we evaluate the feasibility and scalability of the \emph{blockchain deployment service} and \emph{data management services} in uBaaS. Finally, we evaluate the feasibility of \emph{smart contract design services} in uBaaS using a real-world quality tracing use case.

\subsection{Experiment environment}
The uBaaS platform was deployed on an Alibaba Cloud\footnote{\url{https://www.aliyun.com/}} virtual machine (2 vCPUs, 8G RAM, 20GB disk). The adopted blockchain is Ethereum 1.5.9-stable, in which the consensus algorithm is Proof-of-Work (PoW). The selected database for off-chain data storage is MySQL 5.7.17. We set up 100 Alibaba Cloud virtual machines (1 vCPUs, 1G RAM) as the blockchain nodes (the number of nodes varies based on the experiment requirements).

\subsection{Evaluation of Blockchain Deployment Service}


We evaluated the scalability of blockchain deployment service by measuring the deployment time of blockchain with different number of nodes. We set the blockchain type as Ethereum and the difficulty as $0x4000$, and filled in the IP addresses of the nodes for deployment. According to the settings, uBaaS automatically deployed the Ethereum blockchain on the target nodes. 

Fig.~\ref{eDeployment} shows the measurement results for deployment time of Ethereum blockchain using uBaaS with increased number of nodes. The deployment time increased almost linearly, showing good scalability. The experiment results also show that the blockchain deployment service in uBaaS is feasible.

\subsection{Evaluation of Data Management Services}

The main functionality provided by the data management services include: 1) generating an on-chain data registry and an off-chain data table in the conventional database based on the data model built by the developers; 2) writing/reading data to/from the on-chain data registry and off-chain data table. Therefore, we evaluated the data management services in a two-fold way: First, we measured the execution time of creating a data table off-chain in a remote conventional database and generating the corresponding data registry smart contract on-chain; Then, we tested the execution time of writing/reading data to/from the off-chain table and on-chain smart contract respectively. Note that the current version of uBaaS encrypts all the data before storing them on-chain. Thus, the time of writing data and reading data to/from on-chain data registry include data encryption and decryption.

Fig.~\ref{createTable} illustrates the execution time of creating an off-chain data table and generating the associated on-chain data registry with the increased number of data record attributes. The execution time of generating a smart contract on blockchain is much higher than creating a table in the conventional database, due to the Proof of Work (POW) mechanism in Ethereum. In addition, 96.7\% of the on-chain data registry smart contracts are generated within 50 seconds. The on-chain data registry smart contract generation time is fluctuating because block generation time varies around an average value according to the difficulty setting of Ethereum.

Fig.~\ref{writingData} demonstrates the execution time of writing data to an off-chain data table and to the associated on-chain data registry smart contract with increased number of attributes. The execution time increased linear, which shows good scalability. Compared to writing data to the off-chain data table in the conventional database, storing data to the on-chain data registry smart contract is much slower since it is time-consuming to generate a new block and include the transactions in the block. The fluctuation of writing data to the deployed smart contract is also because the block generation time is unstable as mentioned above. 

Fig.~\ref{queryingData} shows the execution time of reading data from the on-chain data registry smart contract, which is much shorter than generating an on-chain data registry smart contract and writing data to the deployed smart contract, as reading is from the local node of blockchain. It is still higher than reading data from traditional database, as the data stored on-chain needs to be decrypted before being presented to the user in uBaaS.

The experiment results in Fig.5-7 also show the feasibility of the data management services in uBaaS.

\subsection{Evaluation of Smart Contract Design Services}
We evaluated the feasibility of the smart contract design services using a real world quality tracing process. Fig.~\ref{qualitytracing} illustrates the quality tracing process for import commodities in China \cite{qualitytracing}. The quality tracing agency accredited by the Chinese government provides quality tracing services and issues traceability certificates of commodity if all requirements are fulfilled. The process starts when a product supplier submits a quality tracing application to the agency. The administrator processes the application paper work (e.g. invoices) and payment. Once the application is validated, the agency assigns a factory inspector to check the factory location, production capability, quality control process, etc. After inspecting the factory, a freight yard inspector is sent to examine the products placed in the freight yard and to inspect the on-site loading process. The inspector seals the containers if the process of on-site loading complies with regulations. In the meantime, a product sample is sent to the lab for sample testing. Once the application passes the inspections and testing, the quality tracing agency issues the supplier a traceability certificate of commodity. In the quality tracing system, the quality tracing agency manages the quality tracing process and data while the lab submits testing reports through the quality tracing system. The quality inspection bureau mainly monitors the quality tracing process and does not provide any input to this system.

\lstset{  
  frame=single,
  framesep=\fboxsep,
  framerule=\fboxrule,
  xleftmargin=\dimexpr\fboxsep+\fboxrule,
  xrightmargin=\dimexpr\fboxsep+\fboxrule,
  language=Java,
  basicstyle=\scriptsize\ttfamily,
  commentstyle=\color{cyan},
  tabsize=2,
  keywordstyle=,
  breaklines=true,  
  captionpos=b,
  escapeinside=``
}

\begin{lstlisting}[caption=Updated \emph{SampleTesting} code after using the \emph{multiple auhtorities} service.,label=MultipleAuthority]
`\textcolor{red}{contract MultipleAuthorities\{}` 
`\textcolor{red}{\quad  uint total;  }` 
`\textcolor{red}{\quad  address[] authority;  }` 
`\textcolor{red}{\quad  bool agreeing;  }` 
`\textcolor{red}{\quad  uint agreeThreshold;  }` 
`\textcolor{red}{\quad  mapping(address => bool) agreeState;  }` 
`\textcolor{red}{\quad  bool agreePermission;}` 
`\textcolor{red}{\quad  address agreeRequester;}` 
`\textcolor{red}{\quad  ...}` 
`\textcolor{red}{\quad  function agreeSignature()\{}` 
`\textcolor{red}{\qquad    agreeState[msg.sender] = true; }` 
`\textcolor{red}{\qquad    if(agreeResult())}` 
`\textcolor{red}{\qquad \quad      agreePermission = true; }` 
`\textcolor{red}{\quad  \}}` 
`\textcolor{red}{\quad  function agreeResult() internal returns (bool signatureResult)\{}` 
`\textcolor{red}{\qquad    uint k = 0; }` 
`\textcolor{red}{\qquad    for(uint i = 0; i <total; i++)}` 
`\textcolor{red}{\quad\qquad      if(agreeState[authority[i]] == true) }` 
`\textcolor{red}{\qquad\qquad        k++;}` 
`\textcolor{red}{\qquad    if(k >= agreeThreshold) }` 
`\textcolor{red}{\qquad\quad      return true; }` 
`\textcolor{red}{\qquad    else }` 
`\textcolor{red}{\qquad\quad      return false; }` 
`\textcolor{red}{\quad  \}}` 
`\textcolor{red}{\quad  function initialAgree() internal\{}` 
`\textcolor{red}{\qquad    ...}` 
`\textcolor{red}{\quad  \}}` 
`\textcolor{red}{\quad  modifier isEnoughAgreement()\{}` 
`\textcolor{red}{\qquad    if(agreeing == true \&\& agreePermission == true \&\& msg.sender == agreeRequester)\{}`  
`\textcolor{red}{\qquad\quad      \underline{\hspace{0.5em}}; }` 
`\textcolor{red}{\qquad      initialAgree(); \}}`  
`\textcolor{red}{\quad  \}}` 
`\textcolor{red}{\quad  ...}` 
`\textcolor{red}{\}}` 

contract SampleTesting `\textcolor{red}{is MultipleAuthorities}`{
  string sampleID;
  bool passed;
  function sampleTest(string ID){
    sampleID = ID;
    passed = false;
  }
  function pass() `\textcolor{red}{isEnoughAgreement()}`{
    passed = true;
  }
  ...
}
\end{lstlisting}

According to the design of the quality tracing system, we deployed an Ethereum blockchain on three nodes, which represent a node in the quality tracing agency, a node in the lab, and a node in the quality inspection bureau. We configured the difficulty as as $0x4000$ and provided the IP addresses of three nodes in uBaaS. Then uBaaS automatically deployed an Ethereum blockchain on the three target nodes.


\begin{lstlisting}[caption=Updated \emph{ServiceAgreement} code after using the \emph{dynamic binding} service.,label=DynamicBinding]
`\textcolor{red}{contract DynamicBinding\{  }`
`\textcolor{red}{\quad  bytes32 hashKey; }`
`\textcolor{red}{\quad  bool init;  }`
`\textcolor{red}{\quad  address owner;}`
`\textcolor{red}{\quad  function initial(bytes32 key)\{}`
`\textcolor{red}{\qquad    if(init != true)\{ }`
`\textcolor{red}{\qquad\quad    hashKey = key; }`
`\textcolor{red}{\qquad\quad      init = true;}`
`\textcolor{red}{\qquad\quad      owner = msg.sender;}`
`\textcolor{red}{\qquad  \}}` 
`\textcolor{red}{\quad \}}`
`\textcolor{red}{\quad  function changeKey(string oldKey , bytes32 newKey)\{}`
`\textcolor{red}{\qquad    if(init == true) }`
`\textcolor{red}{\qquad\quad      if(hashKey == sha256(oldKey))}`
`\textcolor{red}{\qquad\qquad        if(owner == msg.sender) }`
`\textcolor{red}{\qquad\qquad\quad          hashKey = newKey; }`
`\textcolor{red}{\quad  \}}`
`\textcolor{red}{\quad  modifier verify(string inputKey)\{}`
`\textcolor{red}{\qquad    if(hashKey == sha256(inputKey))\{  \underline{\hspace{0.5em}}; \}}` 
`\textcolor{red}{\quad  \}}`
`\textcolor{red}{ \} }`

contract ServiceAgreement `\textcolor{red}{is DynamicBinding}`{
  string firstParty;
  string secondParty;
  bytes32 contractHash;
  ...
  function queryAgreement(`\textcolor{red}{string key}`) 
`\qquad\qquad\textcolor{red}{verify(key)}` constant returns (string, string, bytes32) {
        return firstParty, secondParty, contractHash;
  }
  ...
}
\end{lstlisting}

As discussed in Section 4,  uBaaS uses the smart contract design services to restrict access to the invocation of the functions in the smart contracts. Thus we selected three different functionalities (i.e. sample testing result approval, service agreement query and freight yard picture storage) to apply each of the proposed smart contract design services: \emph{multiple authorities service}, \emph{dynamic binding service}, and \emph{embedded permission service}. We wrote three smart contracts implementing the business logic in those three scenarios and select a function in each of the three smart contract to apply each smart contract design service. Thus, the output of uBaaS is the updated smart contract code with permission control for the smart contract function invocation. List 1-3 show the generated code by uBaaS. The code in black colour represents the input smart contract code (i.e. the smart contract code written by the developers), while the code highlighted in red colour is the code generated by the corresponding smart contract design services in uBaaS.

\begin{lstlisting}[caption=Updated \emph{FreightYardPic} code after using the \emph{embedded permission} service.,label=EmbeddedPermission]
`\textcolor{red}{contract EmbeddedPermission\{}`
`\textcolor{red}{\quad	address [] authority;}`
`\textcolor{red}{\quad	address owner;}`
`\textcolor{red}{\quad	function EmbeddedPermission(address [] temAuthority)\{ }`
`\textcolor{red}{\qquad		owner = msg.sender;}`
`\textcolor{red}{\qquad		authority = temAuthority;}`
`\textcolor{red}{\quad	\}}`
`\textcolor{red}{\quad	function changeAuthority(address [] temAuthority)\{}`
`\textcolor{red}{\qquad		if(msg.sender == owner)\{}`
`\textcolor{red}{\qquad\quad			authority = temAuthority;}`
`\textcolor{red}{\qquad		\}}`
`\textcolor{red}{\quad	\}}`
`\textcolor{red}{\quad	modifier permission()\{}`
`\textcolor{red}{\qquad		for(uint i = 0; i < authority.length; i++)\{}`
`\textcolor{red}{\quad\qquad			if(msg.sender == authority[i])\{}`
`\textcolor{red}{\qquad\qquad				\underline{\hspace{0.5em}};}`
`\textcolor{red}{\qquad\qquad				break;\}\}}`
`\textcolor{red}{\quad	\}}`
`\textcolor{red}{ \}}`

contract FreightYardPic `\textcolor{red}{is EmbeddedPermission}`{ 
	bytes32 [] freightYardPic; 
	address [] freightYardExaminer; 
`\textcolor{red}{\quad	function FreightYardPic() embeddedPermission(addr)\{ }`
`\textcolor{red}{\qquad		address [] addr;}`
`\textcolor{red}{\qquad		addr.push(/authority address/);}`
`\textcolor{red}{\quad	\} }`
	function setFreightYardPic(bytes32 pic, address uploader) `\textcolor{red}{permission()}` { 
		freightYardPic.push(pic); 
        freightYardExaminer.push(uploader);  
	}  
	function getFreightYardPic(uint i) constant returns (bytes32 , address){ 
		return (freightYardPic[i], freightYardExaminer[i]);  
	}  
} 
\end{lstlisting}

The SampleTesting smart contract implements the logic in the sample testing  activity. According to the process design, the product passes the sample test (i.e. the pass() function in the SampleTesting smart contract is invoked) only when enough number of labs agree that the product passes the sample test. Thus, SampleTesting smart contract code and the required lab addresses are provided as the input for uBaaS and the pass() function is selected to apply \emph{multiple authorities} design pattern using the corresponding service. MutlipleAuthorities smart contract including the modifer isEnoughAgreement() attached to pass() are added after using the uBaaS \emph{multiple authorities} service. Listing~\ref{MultipleAuthority} presents the updated smart contract code generated by the \emph{multiple authorities} service in uBaaS.


The quality tracing service agreement signed between the product supplier and the quality tracing service agency is implemented in the smart contract ServiceAgreement. In order to ensure the data privacy in the service agreement, uBaaS adds the modifier verfy() to the queryAgreement() function in ServiceAgreement using uBaaS. Only the party who can provide the secret key can successfully read the service agreement information stored on-chain. The smart contract code generated by the uBaaS \emph{dynamic binding} service is presented in Listing~\ref{DynamicBinding}. Note that the secret key can be changed on demand by invoking function changeKey().\par


The smart contract FreightYardPic maintains the pictures taken at the freight yard. According to the process design, only the freight yard inspectors can upload the hash value of the pictures taken at the freight yard. Thus, we use the \emph{EmbeddedPermission} service in uBaaS to add the access control for the function setFreightYardPic() which only allows the specific agency employee addresses to store the hash value of the freight yard pictures on blockchain. The smart contract code generated by the \emph{embedded permission} service is presented in Listing~\ref{EmbeddedPermission}.

\section{Conclusion and Future Work}
In this paper, we present a unified blockchain as a service platform named uBaaS which provides deployment as a service, design pattern as a service and auxiliary services. Deployment as a service in uBaaS is not vendor locked and provides one-click deployment service by hiding deployment script from users. Design pattern as a service leverage  design patterns to facilitate data management and smart contract design of blockchain-based applications. We evaluate the feasibility and scalability of uBaaS using a real-world quality tracing use case, which demonstrates it is feasible and scalable to design and deploy blockchain-based applications using uBaaS. The future work includes adding more blockchain platforms as deployment platform options, and designing self-sovereign identity as a service in uBaaS.





\section*{References}
\label{references}








\end{document}